# Photometric and spectroscopic observations of the outburst of the symbiotic star AG Draconis between March and June 2016

David Boyd
*Variable Star Section, British Astronomical Association, [davidboyd@orion.me.uk]*

**Abstract**

The symbiotic star AG Dra experienced a double-peaked outburst of 0.6 magnitudes in April and May 2016. Photometry and spectroscopy through the outburst showed the B-V colour index varying linearly with the V magnitude and enabled the temperature variation of the hot star to be calculated from the changing flux in the Hβ and He II 4686Å emission lines.

**Introduction**

The symbiotic star AG Dra is a binary with orbital period 550 days consisting of a hot star, probably a white dwarf, and a cool K-type red giant. These stars are surrounded by a circumbinary nebula generated by a wind from the giant star. This wind is partially ionised by radiation from the hot star creating strong emission lines in the optical spectrum. During quiescence the hot star sustains a high surface temperature by nuclear burning hydrogen-rich material accreted from the stellar wind onto its surface.

**Photometry**

After a long period of relative quiescence at around V magnitude 9.7, the star brightened during April 2016 peaking twice and reaching V = 9.1. Although not as bright an outburst as the one in 2006 which reached V = 8.3, it was nevertheless the brightest it has been recorded for the last 8 years. Figure 1 shows the AAVSO B and V light curves of AG Dra between March 1 and June 14 with crosses marking my B and V magnitude measurements. The B-V colour index varied almost linearly with the V magnitude during the outburst becoming bluer as the system brightened (Figure 2). AG Dra eventually returned to a level about 0.1 mag in V brighter than before the outburst.

**Spectroscopy**

I recorded spectra on 13 dates marked in Figure 1 using a LISA spectrograph (spectral resolution ~1000) attached to a C11 scope. The spectra were flux calibrated using the concurrently measured V magnitudes. Figure 3 presents four spectra recorded during the outburst showing the continuum level rising as the star brightened and changes in the relative strengths of the He II 4686Å and Hβ nebular emission lines.

According to the analysis in Iijima (1981), quoted in Leedjarv et al. (2016), the following relationship between the fluxes of the He II 4686Å, He I 4471Å and Hβ emission lines can be taken as a proxy for the temperature of the hot star which is ionising the giant star's wind.

$$T (hot\ star) = (19.38 * SQRT (2.22*Flux\ 4686 / (4.16*Flux\ H\beta + 9.94*Flux\ 4471)) + 5.13) * 10{,}000\ K$$

In AG Dra, the flux of the He I 4471Å line is sufficiently small that it can be neglected. The fluxes of the He II 4686Å and Hβ emission lines were measured in the flux calibrated spectra and the hot star temperature calculated. Table 1 lists the line fluxes and corresponding hot star temperatures for each spectrum recorded during the outburst. The hot star temperatures are plotted in Figure 4. The temperature errors are derived from a conservative 10% uncertainty in measuring the line fluxes.

This is referred to as a "hot" outburst by González-Riestra et al. (1999). The nature and mechanism of these outbursts is still unclear. The most promising explanation is a combination of increased accretion of hydrogen-rich material from the wind onto the hot star and enhanced nuclear shell burning on its surface. It is still unclear whether AG Dra has an accretion disk.

This is an example of how useful scientific information can be extracted from amateur spectra with careful observing technique, calibration and analysis.

**Acknowledgement**

We acknowledge with thanks the variable star observations from the AAVSO International Database contributed by observers worldwide and used in this research.

**References**

Iijima T., in Photometric and Spectroscopic Binary Systems, ed. E. B. Carling & Z. Kopal, Dordrecht, Kluwer, 517 (1981)
González-Riestra R. et al., Astronomy & Astrophysics, 347, 478 (1999)
Leedjarv L. et al., Mon. Not. R. Astron. Soc., 456, 2558 (2016)

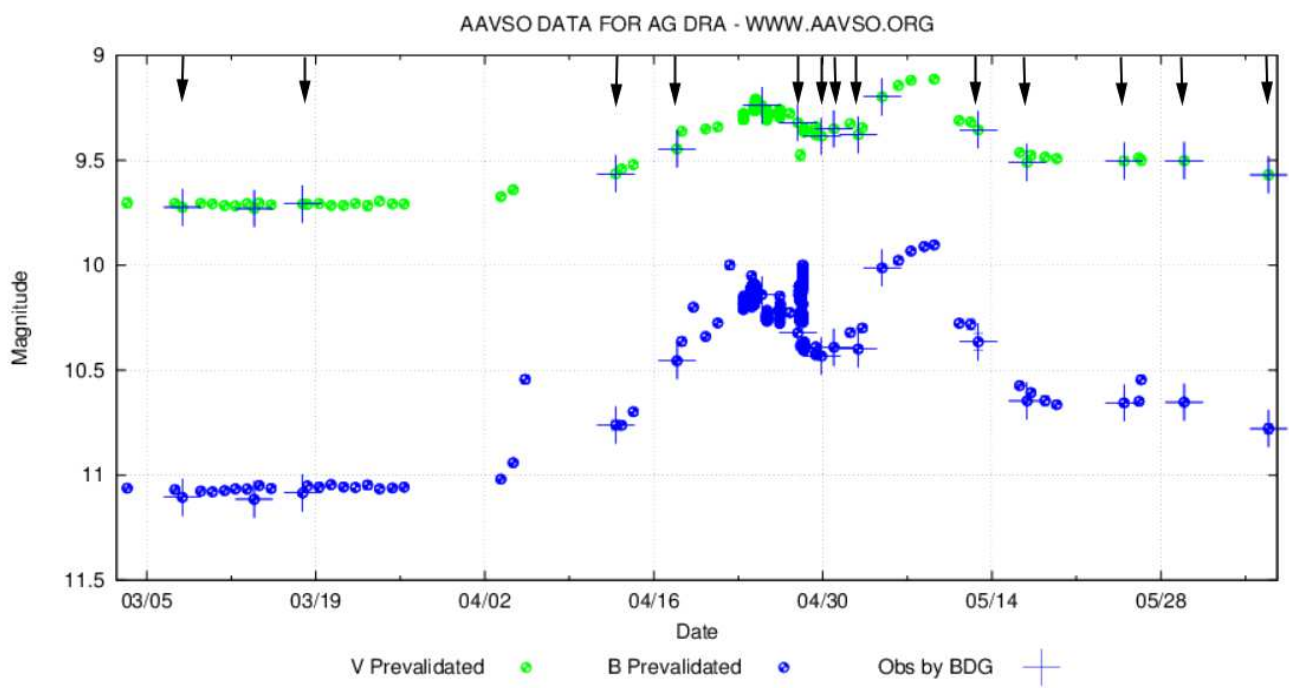

Figure 1. AAVSO B and V-band light curves of AG Dra between March 1 and June 14. Observations by the author are marked with crosses and the arrows at the top indicate the dates on which spectra were recorded.

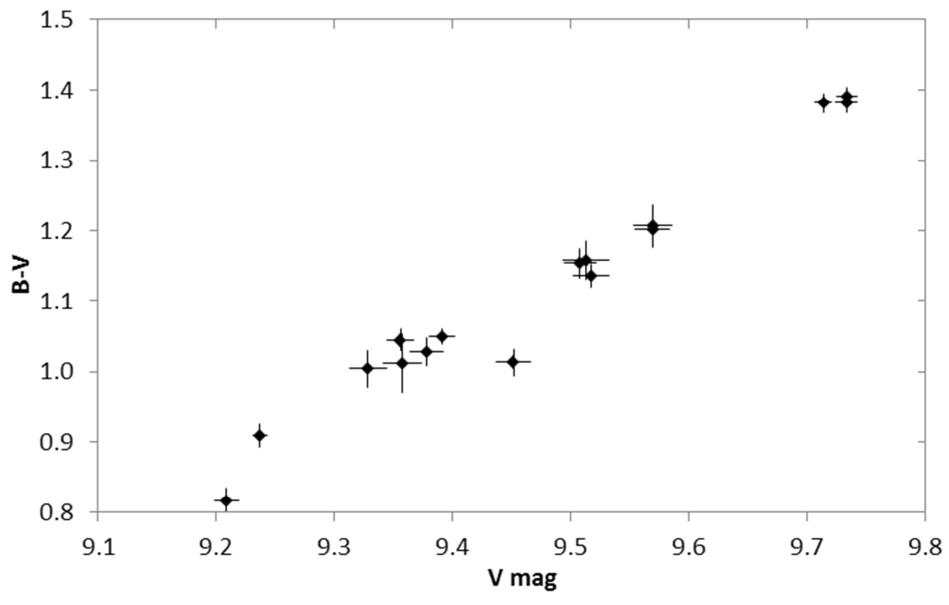

Figure 2. Variation of B-V colour index with V magnitude throughout the outburst.

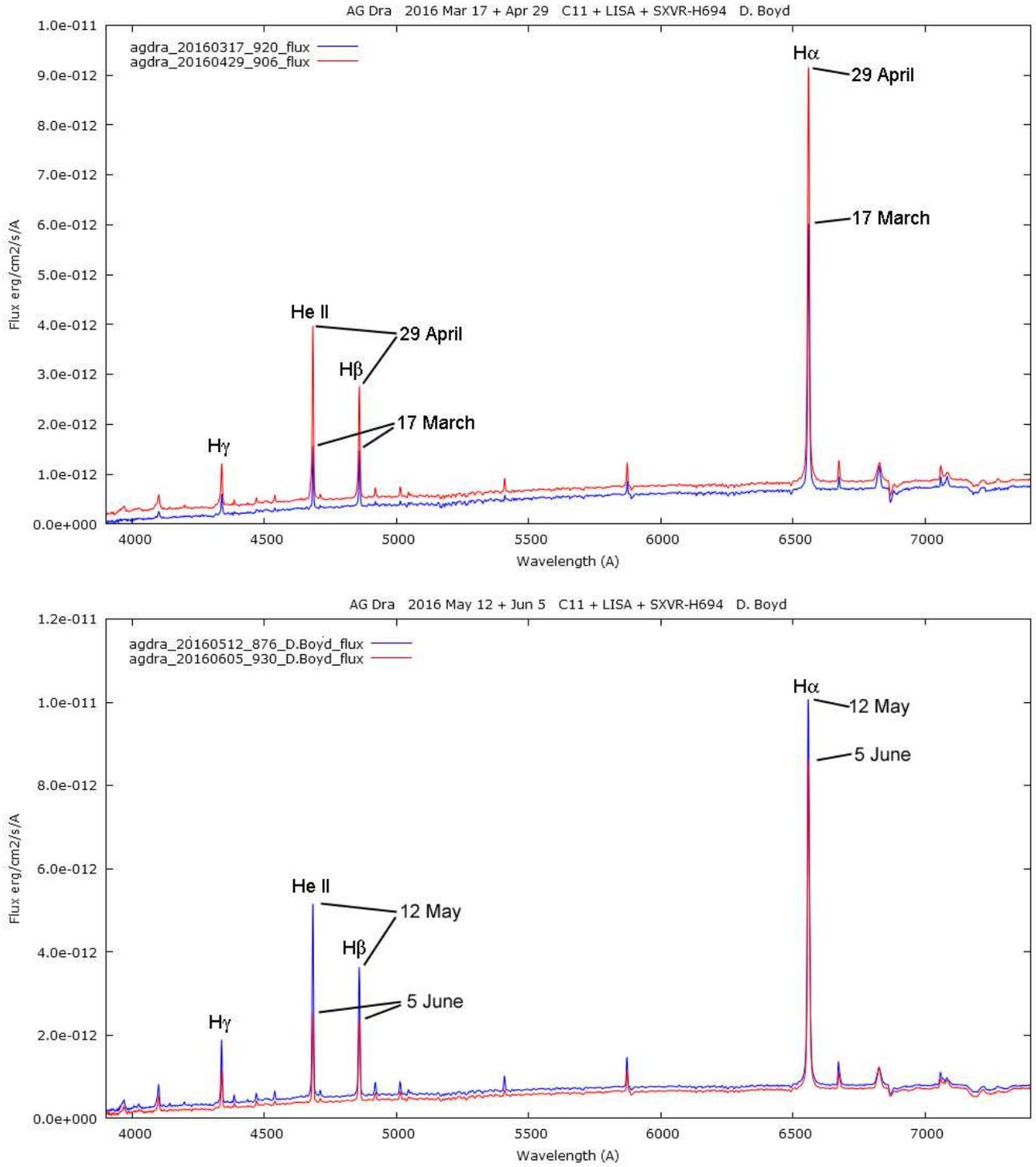

Figure 3. Flux calibrated spectra of AG Dra on March 17, April 29, May 12 and June 5 showing the variation in continuum level and strength of the emission lines.

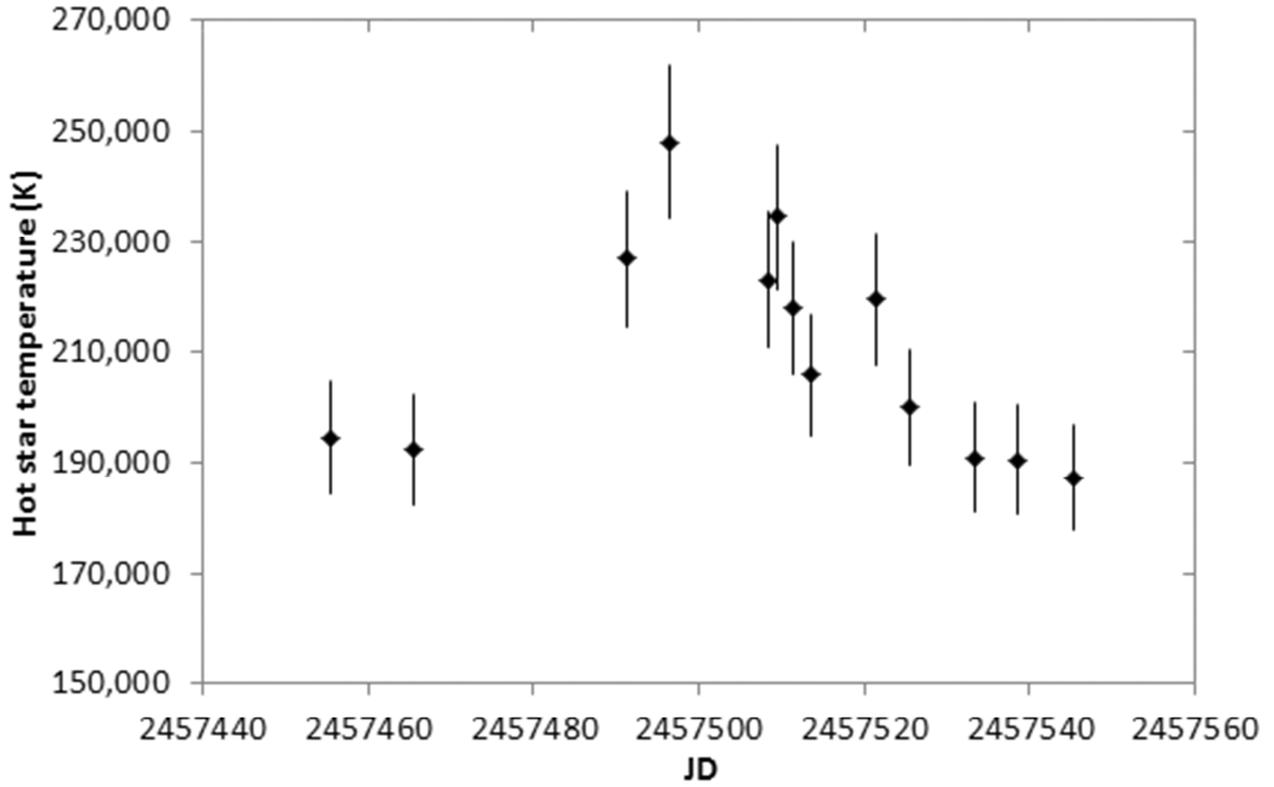

Figure 4. Variation in temperature of the hot star during the outburst.

Table 1. V-band magnitudes, He II 4686Å and Hβ line fluxes and corresponding hot star temperatures.

| Date | JD | V magnitude | He II line flux (erg/cm$^2$/s) | Hβ line flux (erg/cm$^2$/s) | He II / Hβ | Hot star temp (K) |
|---|---|---|---|---|---|---|
| 07-Mar-16 | 2457455 | 9.73 | 7.67E-12 | 7.49E-12 | 1.02 | 194,587 |
| 17-Mar-16 | 2457465 | 9.71 | 7.47E-12 | 7.51E-12 | 0.99 | 192,483 |
| 12-Apr-16 | 2457491 | 9.57 | 1.64E-11 | 1.07E-11 | 1.54 | 226,844 |
| 17-Apr-16 | 2457496 | 9.45 | 2.55E-11 | 1.32E-11 | 1.93 | 248,141 |
| 29-Apr-16 | 2457508 | 9.39 | 2.38E-11 | 1.62E-11 | 1.47 | 223,195 |
| 30-Apr-16 | 2457509 | 9.36 | 2.68E-11 | 1.60E-11 | 1.68 | 234,546 |
| 02-May-16 | 2457511 | 9.38 | 2.52E-11 | 1.82E-11 | 1.39 | 218,034 |
| 04-May-16 | 2457513 | 9.21 | 2.13E-11 | 1.79E-11 | 1.19 | 205,915 |
| 12-May-16 | 2457521 | 9.36 | 3.28E-11 | 2.32E-11 | 1.41 | 219,701 |
| 16-May-16 | 2457525 | 9.52 | 2.21E-11 | 2.00E-11 | 1.10 | 200,087 |
| 24-May-16 | 2457533 | 9.51 | 1.69E-11 | 1.74E-11 | 0.97 | 190,916 |
| 29-May-16 | 2457538 | 9.51 | 1.88E-11 | 1.95E-11 | 0.97 | 190,563 |
| 05-Jun-16 | 2457545 | 9.57 | 1.50E-11 | 1.62E-11 | 0.92 | 187,361 |